\documentstyle[preprint,prc,aps]{revtex}
\begin{document}
\catcode`\@=11
\def\eqalign#1{\null\,\vcenter{\openup\jot\m@th
\ialign{\strut\hfil$\displaystyle{##}$&$\displaystyle{{}##}$\hfil
     \crcr#1\crcr}}\,}
\catcode`\@=12
\title{Positive parity pentaquarks in a Goldstone boson exchange model}
\author{Fl. Stancu\thanks{e-mail : fstancu@ulg.ac.be}}
\address{Universit\'{e} de Li\`ege, Institut de Physique B.5, Sart Tilman,
B-4000 Li\`ege 1, Belgium}
\date{\today}
\maketitle
\everymath={\displaystyle}
\begin{abstract}
We study the stability of the pentaquarks $uudd\overline{Q}$,
$uuds\overline{Q}$ and $udss\overline{Q}$ (Q = c, b, or t) of positive parity
in a constituent quark model based on Goldstone boson exchange interaction
between quarks. The pentaquark parity is the antiquark parity times that of a
quark excited to a p-shell. We show that the Goldstone boson exchange
interaction favors these pentaquarks much more than the negative parity ones of
the same flavour content but all quarks in the ground state. We find that the
nonstrange pentaquarks are stable against strong decays.
\end{abstract}

\vspace{1cm}

The existence of particles made of more than three quarks is an important issue
of QCD inspired model. The interest has been raised so far by particles
described by the colour state ${\left[{222}\right]}_{C}$, the tetraquarks
$q^2\overline{q}^2$, the pentaquarks $q^4\overline{q}$ and the hexaquarks $q^6$.
The present study is devoted to pentaquarks, first proposed independently by
Gignoux, Silvestre-Brac and Richard \cite{GI87} and Lipkin \cite{LI87} about
ten years ago. Within a constituent quark model based on one-gluon exchange
(OGE) interaction, these authors found that the states
${P}_{\overline{c}s}^{0}\ =\
\left.{\left|{uuds\overline{c}}\right.}\right\rangle$ and
${P}_{\overline{c}s}^{-}\ =\
\left.{\left|{udds\overline{c}}\right.}\right\rangle$ and their conjugates are
stable against strong decays. Within better approximations, they turned out to
be unstable \cite{FL89,ZO94}. A systematic theoretical study \cite{LE89} in a
model with OGE interaction suggested several candidates for stability, and
among them especially those with strangeness S = -1 or -2. In particular, the
$uuds\overline{c}$ system was bound by -52 MeV. The nonstrange systems
$uudd\overline{Q}$ (Q = c or b) were unbound.\par
Calculations done within an instanton model \cite{TA93} or a Skyrme model
\cite{RI93} show that pentaquarks, irrespective of their strangeness, appear as
bound or near-threshold resonances depending on the model parameters. Moreover,
the lowest pentaquark states predicted in Ref. \cite{RI93} have positive parity
(L = 1).\par
If bound, the lifetime of the pentaquark $uuds\overline{c}$ or $udds\overline{c}$
is expected to be similar to that of the $D_s^{\pm}$ meson. Using various
mechanisms, the typically estimated pentaquark production cross section is of
the order of $1\ \%$ that of $D_s^{\pm}$ \cite{MO96}. Based on the above
theoretical predictions, experiments are being planned and the first search for
the pentaquarks $P_{\overline{c}s}^0$ and
$P_{\overline{c}s}^-$, performed at Fermilab, has just been
reported \cite{AI97}. The decay ${P}_{\overline{c}s}^{0}\ \rightarrow \
\phi \pi
 p$ was analyzed for two different pentaquark masses : M = 2.75 GeV, being
the lowest mass expected from the OGE model and M = 2.86 GeV, the value at
which the largest number of events was observed. No convincing evidence for
pentaquarks decaying to $\phi\pi p$ was observed yet.\par
The theoretical predictions are definitely model-dependent. In a previous work
\cite{GE98}, we studied the stability of heavy-flavoured pentaquarks within a
chiral constituent quark model \cite{GL96a,GL96b,GL97a} originally proposed by
Glozman and Riska \cite{GL96a}. In this model, the hyperfine splitting in
hadrons is obtained from the short-range part of the Goldstone boson exchange
(GBE) interaction between quarks, instead of the OGE interaction of
conventional models, as discussed above. The main merit of the GBE interaction
is that it reproduces the correct ordering of positive and negative parity
states in all parts of the considered spectrum \cite{GL96b,GL97a,GL98} in
contrast to any other OGE model.\par
In Ref. \cite{GE98}, we considered pentaquarks with strangeness ranging from S
= -3 to S = 0. There, we assumed that all light quarks are identical and the
ground-state orbital (O) wave function is symmetric under permutation of light
quarks, i.e. it corresponds to the Young diagram ${\left[{4}\right]}_{O}$. The
subsystem of light quarks must necessarily be in a colour 3, or alternatively a
${\left[{211}\right]}_{C}$ state. Then, the Pauli principle allows a certain
number of spin ${\left[{f}\right]}_{S}$ and flavour ${\left[{f}\right]}_{F}$
states to be combined with ${\left[{4}\right]}_{O}$ and
${\left[{211}\right]}_{C}$ to give a total antisymmetric four-quark state. We
found that any of these states combined with a heavy antiquark $\overline{Q}$
where Q = c or b gave rise to a pentaquark energy which was at least 300-400
MeV above the dissociation threshold nucleon plus meson, i.e. the considered
pentaquark cannot be a bound compact object. Its parity is P = -1, due to the
antiquark.\par
The novelty of this study is that, within the same GBE model, we analyse the
stability of pentaquarks with P = +1. In such a case, the parity of the
pentaquark is given by P$\ =\ {\left({-}\right)}^{\mbox{L\ +\ 1}}$, thus the light
quarks must carry an angular momentum L odd. Here, we consider the case L = 1,
which implies that the subsystem of four light quarks must be in a state of
orbital symmetry ${\left[{31}\right]}_{O}$. Although the kinetic energy of such
a state is higher than that of the totally symmetric ${\left[{4}\right]}_{O}$
state, a schematic estimate \cite{GL98} suggests that the
${\left[{31}\right]}_{O}$ symmetry would lead to a stable pentaquark. In the
following, we give the arguments of Ref. \cite{GL98} based on a simplified GBE
interaction of the form
\begin{equation}
{V}_{\chi }\ =\ -\ {C}_{\chi }\ \sum\limits_{ i\ <\ j}^{} {\lambda }_{
i}^{ F}.{\lambda }_{ j}^{ F} \ {\vec{\sigma }}_{i}.{\vec{\sigma
}}_{j}
\end{equation}
with $\lambda_i^F$ the Gell-Mann matrices, $\vec{\sigma}_i$ the Pauli matrices
and $C_{\chi}\cong$ 30 MeV, determined from the $\Delta$-N splitting
\cite{GL96a}.\par
In the spirit of Glozman and Riska's model, there is no meson-exchange
interaction between quarks and antiquarks. It is assumed that the
$q\overline{q}$ pseudoscalar pair interaction is automatically included in the
GBE interaction \cite{GL96a}. Then, as far as the spin-spin interaction is
concerned, the discussion is restricted to the light $q^4$ subsystem. The Pauli
principle allows for the following two totally antisymmetric states with
${\left[{31}\right]}_{O}$ symmetry, written in the flavour-spin (FS) coupling
scheme
\begin{equation}
\left.{\left|{1}\right.}\right\rangle\ =\
\left({{\left[{31}\right]}_{O}{\left[{211}\right]}_{C}{\left[{{1}^{4}}\right]}_{
OC}\
;\
{\left[{22}\right]}_{F}{\left[{22}\right]}_{S}{\left[{4}\right]}_{FS}}\right)
\end{equation}
\begin{equation}
\left.{\left|{2}\right.}\right\rangle\ =\
\left({{\left[{31}\right]}_{O}{\left[{211}\right]}_{C}{\left[{{1}^{4}}\right]}_{
OC}\
;\
{\left[{31}\right]}_{F}{\left[{31}\right]}_{S}{\left[{4}\right]}_{FS}}\right)
\end{equation}
Asymptotically, a ground state baryon and a meson, into which a pentaquark can
split, would give ${\left[{3}\right]}_{O}\ \times  \ {\left[{2}\right]}_{O}\
=\ {\left[{5}\right]}_{O}\ +\ {\left[{41}\right]}_{O}\ +\
{\left[{32}\right]}_{O}$. By removing the antiquark, one can make the reduction
${\left[{41}\right]}_{O}\ \rightarrow  \ {\left[{31}\right]}_{O}\ \times
\ {\left[{1}\right]}_{O}$ or ${\left[{32}\right]}_{O}\ \rightarrow  \
{\left[{31}\right]}_{O}\ \times
\ {\left[{1}\right]}_{O}$. Thus, the symmetry ${\left[{31}\right]}_{O}$ of the
light quark subsystem is compatible with an L = 1 asymptotically separated
baryon plus meson system.\par
The expectation value of (1) calculated, for example, according to the Appendix
of Ref. \cite{ST97}, is $-28 \ C_{\chi}$ for
$\left|{\left.{1}\right\rangle}\right.$ and $-64/3 \ C_{\chi}$ for
$\left|{\left.{2}\right\rangle}\right.$. These two states would actually couple
via a quark-antiquark spin-spin interaction to a total angular momentum J$ =\
\frac{1}{2}$ or $\frac{3}{2}$ where $\vec{\mbox{J}}\ =\ \vec{\mbox{L}}\ +\ \vec{\mbox{S}}\ +\
{\vec{\mbox{s}}}_{Q}$, with \, $\vec{\mbox{L}}$, $\vec{\mbox{S}}$ \, the 
angular momentum and spin of the
light system and $\vec{\mbox{s}}_Q$ the spin of the antiquark. As 
the quark-antiquark
interaction is neglected here, in the following we restrict our discussion to
the lowest state, i.e. $\left|{\left.{1}\right\rangle}\right.$. The
quark-antiquark interaction is neglected in the description of mesons as well,
as for example in Ref. \cite{PE97}, so that the meson Hamiltonian contains a
kinetic and a confinement term only.\par
We are interested in the quantity
\begin{equation}
\Delta E =\ E(q^4\overline{Q}) -\ E(q^3) -\
E(q\overline{Q})
\end{equation}
In our schematic estimate, we suppose that the confinement energy roughly
cancels out in $\Delta E$. Then, the kinetic energy contribution \, 
$\Delta KE$ \, to \,
$\Delta E$ \, is \, $\Delta  KE\ =\ 5/4\ \hbar \omega$ in a harmonic 
oscillator model
and the GBE contribution for the state $\left|{\left.{1}\right\rangle}\right.$
is $\Delta V_{\chi} =\ - 14 \ C_{\chi}$. With $\hbar\omega \approx$ 250 MeV,
determined from the N(1440) \, - \, N splitting \cite{GL96a}, this would give
\begin{equation}
\Delta  E\ =\ {\frac{5}{4}}\ \hbar \omega  \ -\ 14\ {C}_{\chi }\ =\ -\
107.5\ MeV
\end{equation}
i.e. a substantial binding. This is to be contrasted with the negative parity
pentaquarks studied in Ref. \cite{GE98} where for the lowest state one has
$\Delta  E\ =\ 3/4\ \hbar \omega  \ -\ 2\ {C}_{\chi }\ =\ 127.5\ MeV$,
i.e. unstability, consistent with the detailed study made in \cite{GE98}.\par
The estimate (5) is a consequence of the flavour dependence of the GBE
interaction. For a specific spin state ${\left[{f}\right]}_{S}$, a schematic
OGE interaction of type ${V}_{c\ m}\ =\ -\ {C}_{c\ m}\ \sum \ {\lambda
}_{i}^{c}.{\lambda }_{j}^{c}\ {\vec{\sigma }}_{i}.{\vec{\sigma }}_{j}$ does not
make a distinction between ${\left[{4}\right]}_{O}$ and
${\left[{31}\right]}_{O}$ so that the ${\left[{31}\right]}_{O}$ state will
appear higher than ${\left[{4}\right]}_{O}$ due to the kinetic energy. The GBE
interaction overcomes the excess of kinetic contribution in
${\left[{31}\right]}_{O}$ and generates a lower expectation value for
${\left[{31}\right]}_{O}$ than for ${\left[{4}\right]}_{O}$.\par
The GBE Hamiltonian has the form \cite{GL96b} :
\begin{equation}
H= \sum_i m_i + \sum_i \frac{\vec{p}_{i}^{\,2}}{2m_i} - \frac {(\sum_i
\vec{p}_{i})^2}{2\sum_i m_i} + \sum_{i<j} V_{\text{conf}}(r_{ij}) + \sum_{i<j}
V_\chi(r_{ij}) \, ,
\label{ham}
\end{equation}
with the linear confining interaction :
\begin{equation}
 V_{\text{conf}}(r_{ij}) = -\frac{3}{8}\lambda_{i}^{c}\cdot\lambda_{j}^{c} \, C
\, r_{ij} \, ,
\label{conf}
\end{equation}
and the spin--spin component of the GBE interaction in its $SU_F(3)$ form :
\begin{eqnarray}
V_\chi(r_{ij})
&=&
\left\{\sum_{F=1}^3 V_{\pi}(r_{ij}) \lambda_i^F \lambda_j^F \right.
\nonumber \\
&+& \left. \sum_{F=4}^7 V_{K}(r_{ij}) \lambda_i^F \lambda_j^F
+V_{\eta}(r_{ij}) \lambda_i^8 \lambda_j^8
+V_{\eta^{\prime}}(r_{ij}) \lambda_i^0 \lambda_j^0\right\}
\vec\sigma_i\cdot\vec\sigma_j,
\label{VCHI}
\end{eqnarray}
\noindent
with $\lambda^0 = \sqrt{2/3}~{\bf 1}$, where $\bf 1$ is the $3\times3$ unit
matrix. The interaction (8) contains $\gamma = \pi, K, \eta$ and $\eta '$
meson-exchange terms and the form of $V_{\gamma} \left(r_{ij}\right)$ is given
as the sum of two distinct contributions : a Yukawa-type potential containing
the mass of the exchanged meson and a short-range contribution of opposite
sign, the role of which is crucial in baryon spectroscopy. For a given meson
$\gamma = \pi, K, \eta$ or $\eta '$, the meson exchange potential is
\begin{equation}V_\gamma (r)=
\frac{g_\gamma^2}{4\pi}\frac{1}{12m_i m_j}
\{\theta(r-r_0)\mu_\gamma^2\frac{e^{-\mu_\gamma r}}{ r}- \frac {4}{\sqrt {\pi}}
\alpha^3 \exp(-\alpha^2(r-r_0)^2)\}
\end{equation}
For the Hamiltonian (6)-(9), we use the
parameters of Refs.\cite{GL96b,GL97a}. These are :
$$\frac{g_{\pi q}^2}{4\pi} = \frac{g_{\eta q}^2}{4\pi} =
\frac{g_{Kq}^2}{4\pi}= 0.67,\,\,
\frac{g_{\eta ' q}^2}{4\pi} = 1.206 , $$
$$r_0 = 0.43 \, { fm}, ~\alpha = 2.91 \, { fm}^{-1},~~
 C= 0.474 \, { fm}^{-2}, \, m_{u,d} = 340 \, { MeV}, \, m_s = 440 \, {MeV}$$
\begin{equation}
 \mu_{\pi} = 139 \, { MeV},~ \mu_{\eta} = 547 \, { MeV},~
\mu_{\eta'} = 958 \, { MeV},~ \mu_{K} = 495 \, { MeV}.
\end{equation}
\par
The masses of the threshold hadrons are calculated variationally as in Ref.
\cite{GE98} where we assume an $s^3$ configuration for baryons. They are given
in Table 1 where a theoretical $\overline{t}s$ meson is also included in order
to analyse the large $m_Q$ limit. The mass of the $\overline{t}s$ meson has
been determined by taking $m_t$ = 175 GeV.\par
For pentaquarks, we used the internal Jacobi coordinates
\begin{eqnarray}
\begin{array}{c}\vec{x}\ =\ {\vec{r}}_{1}\ -\ {\vec{r}}_{2}\ , \, \hspace{5mm} \vec{y}\ =\
{\left({{\vec{r}}_{1}\ +\ {\vec{r}}_{2}\ -\ 2{\vec{r}}_{3}}\right)/\sqrt
{3}}\\
\\
\vec{z}\ =\ {\left({{\vec{r}}_{1}\ +\ {\vec{r}}_{2}\ +\ {\vec{r}}_{3}\ -\
3{\vec{r}}_{4}}\right)/\sqrt {6}}\ , \, \hspace{5mm} \vec{t}\ =\
{\left({{\vec{r}}_{1}\
+\ {\vec{r}}_{2}\ +\ {\vec{r}}_{3}+\ {\vec{r}}_{4}-\
4{\vec{r}}_{5}}\right)/\sqrt {10}}\end{array}
\end{eqnarray}
First, we expressed the $q^4$ orbital wave functions of symmetry
${\left[{31}\right]}_{O}$ in terms of the above Jacobi coordinates. We assumed
an $s^3p$ structure for ${\left[{31}\right]}_{O}$ and inspired by Moshinski's
method \cite{MO69}, we found the content of the three independent
${\left[{31}\right]}_O$ states \cite{ST96}, denoted below by $\psi_i$, in
terms of shell model functions $\left.{\left|{n\ \ell \
m}\right.}\right\rangle$. The result is \\


\begin{eqnarray}
{\psi }_{1} = \renewcommand{\arraystretch}{0.5}
\begin{array}{c} $\fbox{1}\fbox{2}\fbox{3}$ \\
$\fbox{4}$\hspace{9mm} \end{array}
=
\left\langle{\vec{x}\left|{000}\right.}\right
\rangle\left\langle{\vec{y}\left|{000}\right.}\right\rangle\left
\langle{\vec{z}\left|{001}\right.}\right\rangle
\end{eqnarray}

\begin{eqnarray}
{\psi }_{2} = \renewcommand{\arraystretch}{0.5}
\begin{array}{c} $\fbox{1}\fbox{2}\fbox{4}$ \\
$\fbox{3}$\hspace{9mm} \end{array}
=
\left\langle{\vec{x}\left|{000}\right.}\right
\rangle\left\langle{\vec{y}\left|{010}\right.}\right\rangle\left
\langle{\vec{z}\left|{000}\right.}\right\rangle
\end{eqnarray}

\begin{eqnarray}
{\psi }_{3} = \renewcommand{\arraystretch}{0.5}
\begin{array}{c} $\fbox{1}\fbox{3}\fbox{4}$ \\
$\fbox{2}$\hspace{9mm} \end{array}
=
\left\langle{\vec{x}\left|{010}\right.}\right
\rangle\left\langle{\vec{y}\left|{000}\right.}\right\rangle\left
\langle{\vec{z}\left|{000}\right.}\right\rangle
\end{eqnarray}

\noindent where, for convenience, we took the quantum number m = 0 everywhere. The
pentaquark orbital wave functions $\psi_i^5$ are then obtained by multiplying
each $\psi_i$ by the wave function
$\left\langle{\vec{t}\left|{000}\right.}\right\rangle$ which describes the
relative motion of the $q^4$ subsystem and the antiquark $\overline{Q}$.
Assuming two variational parameters, $a$ for the internal motion of $q^4$ and
$b$ for the relative motion of $q^4$ and $\overline{Q}$, we have explicitly
\begin{equation}
{\psi }_{1}^{5}\ =\ N\ \exp\ \left[{-\ {\frac{a}{2}}\ \left({{x}^{2}\ +\
{y}^{2}\ +\ {z}^{2}}\right)\ -\ {\frac{b}{2}}\ {t}^{2}}\right]\ z\ {Y}_{10}\
\left({\hat{z}}\right)
\end{equation}
\begin{equation}
{\psi }_{2}^{5}\ =\ N\ \exp\ \left[{-\ {\frac{a}{2}}\ \left({{x}^{2}\ +\
{y}^{2}\ +\ {z}^{2}}\right)\ -\ {\frac{b}{2}}\ {t}^{2}}\right]\ y\ {Y}_{10}\
\left({\hat{y}}\right)
\end{equation}
\begin{equation}
{\psi }_{3}^{5}\ =\ N\ \exp\ \left[{-\ {\frac{a}{2}}\ \left({{x}^{2}\ +\
{y}^{2}\ +\ {z}^{2}}\right)\ -\ {\frac{b}{2}}\ {t}^{2}}\right]\ x\ {Y}_{10}\
\left({\hat{x}}\right)
\end{equation}
where
\begin{equation}
N\ =\ {\frac{{2}^{3/2}{a}^{11/4}{b}^{3/4}}{{3}^{1/2}{\pi }^{5/2}}}
\end{equation}
The kinetic energy part of (6) can be calculated analytically.  For the state
(2) or (3), its form is
\begin{eqnarray}
\begin{array}{lcl}\left\langle{T}\right\rangle\ &=&\ {\frac{1}{3}}\
\left[{\left\langle{{\psi }_{1}^{5}\left|{T}\right|{\psi
}_{1}^{5}}\right\rangle\ +\ \left\langle{{\psi }_{2}^{5}\left|{T}\right|{\psi
}_{2}^{5}}\right\rangle\ +\ \left\langle{{\psi }_{3}^{5}\left|{T}\right|{\psi
}_{3}^{5}}\right\rangle}\right]\\
\\ &=&\ {\hbar }^{2}\ \left({{\frac{11}{2{\mu }_{1}}}\ a\ +\ {\frac{3}{{2\mu
}_{2}}}\ b}\right)\end{array}
\end{eqnarray}
with
\begin{eqnarray}
{\frac{4}{{\mu }_{1}}}\ =\ \left\{{\ \begin{array}{c}{\frac{1}{{m}_{1}}}\ +\
{\frac{3}{{m}_{2}}}\ \ for\ \ {q}_{1}{q}_{2}^{3}\\
\\ {\frac{2}{{m}_{1}}}\ +\ {\frac{2}{{m}_{2}}}\ \ for\ \
{q}_{1}^{2}{q}_{2}^{2}\end{array}}\right.
\end{eqnarray}
where $q_1$, $q_2$ are light quarks and
\begin{equation}
{\frac{5}{{\mu }_{2}}}\ =\ {\frac{1}{{\mu }_{1}}}\ +\ {\frac{4}{{m}_{Q}}}
\end{equation}
$m_Q$ representing the heavy antiquark mass. Here, we choose $m_c$ = 1.35 GeV,
$m_b$ = 4.66 GeV according to Ref. \cite{GE98} and $m_t$ = 175 GeV. Taking
$m_u$ = $m_d$ = $m_s$ = $m_Q$ and setting a=b, the identical particle limit
$\left\langle{T}\right\rangle\ =\ {\frac{7}{2}}\ \hbar \omega$ with
$\hbar \omega  \ =\ 2\ a{\hbar }^{2}/m$ is recovered correctly.\par
Integrating in the coulour space as shown in Ref. \cite{GE98}, the confinement
part of (6) becomes
\begin{equation}
\left\langle{{V}_{conf}}\right\rangle\ =\ {\frac{C}{2}}\ \left({6\
\left\langle{{r}_{12}}\right\rangle\ +\ 4\
\left\langle{{r}_{45}}\right\rangle}\right)
\end{equation}
where the coefficients 6 and 4 account for the number of light-light and
light-heavy pairs, respectively, and
\begin{equation}
\left\langle{{r}_{ij}}\right\rangle\ =\ {\frac{1}{3}}\
\left[{\left\langle{{\psi }_{1}^{5}\left|{{r}_{ij}}\right|{\psi
}_{1}^{5}}\right\rangle\ +\ \left\langle{{\psi
}_{2}^{5}\left|{{r}_{ij}}\right|{\psi }_{2}^{5}}\right\rangle\ +\
\left\langle{{\psi }_{3}^{5}\left|{{r}_{ij}}\right|{\psi
}_{3}^{5}}\right\rangle}\right]
\end{equation}
An analytic evaluation gives
\begin{equation}
\left\langle{{r}_{12}}\right\rangle\ =\ {\frac{20}{9}}\ \sqrt {{\frac{1}{\pi
 a}}}
\end{equation}
and
\begin{equation}
\left\langle{{r}_{45}}\right\rangle\ =\ {\frac{1}{3\sqrt {2\pi }}}\
\left[{2\sqrt {{\frac{3}{a}}\ +\ {\frac{5}{b}}}\ +\ \sqrt {5b}\
\left({{\frac{1}{2a}}\ +\ {\frac{1}{b}}}\right)}\right]
\end{equation}
\par
The matrix elements of the spin-flavour operators of (8) have been calculated
using the fractional parentage technique described in Ref. \cite{ST96} based on
Clebsch-Gordan coefficients of the group $S_4$ \cite{PE96}. In this way, each
matrix element reduces to a linear combination of two-body matrix elements of
either symmetric or antisymmetric states for which Eqs. (3.3) of Ref.
\cite{GL96a} can be used to integrate in the spin-flavour space. The resulting
linear combinations contain orbital two-body matrix elements of type
$\left\langle{ss\left|{{V}_{\gamma }}\right|ss}\right\rangle\ ,\
\left\langle{sp\left|{{V}_{\gamma }}\right|sp}\right\rangle$ and
$\left\langle{sp\left|{{V}_{\gamma }}\right|ps}\right\rangle$ where $\gamma =
\pi, K, \eta$ and $\eta '$ as in Eq. (8).\par
In Table 2, we present results for S = 0, S = -1 and S = -2 pentaquarks. The
quantity $\Delta E$, defined by (4) and exhibited in column 5, is obtained from
$E\left(q^4\overline{Q}\right)$ calculated variationally for the state
$\left.{\left|{1}\right.}\right\rangle$ defined by (2). The optimal values of
the parameters $a$ and $b$ are indicated in each case. In all cases, one has
$a\ >\ b$. The inverses 1/a and 1/b give an idea of the quark-quark and
quark-antiquark distances, respectively. One can see that at equilibrium the
light quarks are clustered together, orbitting around the heavy antiquark.\par
The present variational solution does not give binding for $uuds\overline{Q}$
and $udss\overline{Q}$. The quantity $\Delta E$ decreases smoothly with
increasing $m_Q$ but does not become negative even for $m_t$=175 GeV.
However, as expected from the discussion around Eq. (5), there is much less
``repulsive" effect in these type of pentaquarks than that appearing for the
negative parity pentaquarks studied in Ref. \cite{GE98}, where $\Delta E \sim$
370-490 MeV for pentaquarks with similar flavour content.\par
But the nonstrange positive parity pentaquarks $uudd\overline{Q}$ are bound by
-75.6 MeV, -95.6 MeV and -102.8 MeV for Q = c, b and t, respectively. The reason is that
the GBE interaction is stronger in the nonstrange case because of the
$1/m_im_j$ factor in Eq. (9). Thus, the GBE model suggests that the nonstrange
positive parity pentaquarks are the best candidates for stable compact systems.
This in contrast to OGE based models where strangeness is required  \cite{LE89}
in order to reach stability for heavy-flavoured pentaquarks. The present
results have similarities with those obtained in \cite{RI93} from the Skyrme
model : 1) the lowest pentaquark states have positive parity for any flavour
content ; 2) stability does not necessarily require strangeness.\par
From Table 2 we can obtain upper limits for the masses of the stable quarks as
$M(uudd\overline{c})$ = 2.895 GeV and
$M(uudd\overline{b})$ = 6.176 GeV.\par
It would be very useful indeed to look for a better variational solution for
positive parity pentaquarks to see whether the soft repulsion persists or
disappears for strange systems and to improve the upper bounds for the masses
of the stable states. A possible way would be to consider interference effects
with other asymptotic channels, as for example, in Ref. \cite{FL89}. It would
be also interesting to see how the binding energy of the pentaquarks depends on
the particular parametrization of the hyperfine interaction (nonrelativistic
\cite{GL96b,GL97a} versus semi-relativistic \cite{GL97b} GBE model).\par
Irrespective of the type of hyperfine interaction, an important issue remains
the model of confinement. Both in OGE based calculations
\cite{LI87,FL89,ZO94,LE89} and here, the confinement relies on a
$\lambda_i^c.\lambda_j^c$ ansatz, which does not respect local colour gauge
invariance. Despite this defect, the first estimates have been done under this
assumption, for simplicity. For baryons, the ansatz $\lambda_i^c.\lambda_j^c$ is
a good approximation and it remains to be seen how good it is for
pentaquarks.\par
But, at any level of accuracy, it is expected that the upper bound obtained for
pentaquarks of negative parity, as those studied in Ref. \cite{GE98}, would
remain higher than the upper bound obtained for positive parity pentaquarks as
introduced here. In an OGE based model, the situation will be the other way
round. It remains to future experiments to disentangle between GBE and OGE
models, through the search of pentaquarks.

\acknowledgements
I am most grateful to Leonid Glozman for several useful comments, and to Dan
Olof Riska for pointing out some features of the Skyrme model results.

\renewcommand{\arraystretch}{1.5}
\begin{table}
\parbox{15cm}{\caption[hadron]{\label{hadron} Masses of hadrons 
required to calculate the
threshold energy
$E_T = E_{baryon} + E_{meson}$. The experimental mass for mesons represents the
average $\overline{M} = \left(M + 3M^{\ast}\right)/4$, when both the
pseudoscalar $M$ and the vector $M^{\ast}$ mesons masses are 
available \cite{PD96} .}}
\begin{tabular}{c|cc}
Hadron \hspace{1cm} & \multicolumn{2}{c}{Mass (GeV)} \\
&variational & experiment\\
\hline
N & 0.970 & 0.939 \\
$\Sigma$ & 1.235 & 1.192 \\
$\overline{D}$ & 2.001 & 1.973 \\
$\overline{D}_s$ & 2.087 & 2.076 \\
$\overline{B}$ & 5.302 & 5.313 \\
$\overline{B}_s$ & 5.379 & 5.375 \\
$\overline{t}s$ & 175.711
\end{tabular}
\end{table}

\begin{table}
\caption{Lowest positive parity pentaquarks of
total angular momentum J$ = 1/2, 3/2$ and symmetry structure given by Eq. (2).
Column 1 gives the flavour content, column 2 the isospin I, columns 3 and
4 the
optimal variational parameters associated to the wave functions (15)-(18),
column 5 gives $\Delta E = E - E_T$ where $E$ is the upper bound for the
expectation value of the Hamiltonian (6)-(10) and $E_T = E_{baryon} +
E_{meson}$ (see Table 1) for the threshold given in the last column.}
\begin{tabular}{cccccc}
Pentaquark & I & \multicolumn{2}{c}{Variational parameters
$(GeV^2)$} &
$\Delta E$  & Threshold \\
& & \hspace{6mm} a & b & (MeV) & \\
\hline
$uudd\overline{c}$ & 0 & \hspace{6mm} 0.110 & 0.040 & -75.6 & $N + \overline{D}$ \\
$uudd\overline{b}$ & 0 & \hspace{6mm} 0.110 & 0.053 & -95.6 & $N + \overline{B}$ \\
$uudd\overline{t}$ & 0 & \hspace{6mm} 0.110 & 0.061 & -102.8 & $N + \overline{t}s$ \\
$uuds\overline{c}$ & 1/2 & \hspace{6mm} 0.101 & 0.041 & 104.7 & $N + \overline{D}_s$ \\
$uuds\overline{b}$ & 1/2 & \hspace{6mm} 0.102 & 0.054 & 93.4 & $N + \overline{B}_s$ \\
$uuds\overline{t}$ & 1/2 & \hspace{6mm} 0.102 & 0.063 & 86.9 & $N + \overline{t}s$ \\
$udss\overline{c}$ & 1 & \hspace{6mm} 0.092 & 0.041 & 81.0 & $\Sigma + \overline{D}_s$ \\
$udss\overline{b}$ & 1 & \hspace{6mm} 0.092 & 0.055 & 69.3 & $\Sigma + \overline{B}_s$ \\
$udss\overline{t}$ & 1 & \hspace{6mm} 0.093 & 0.064 & 62.6 & $\Sigma + \overline{t}s$
\end{tabular}
\end{table}

\end{document}